\documentclass{PoS}

\usepackage{amsmath}

\title{TKNN formula for general lattice Hamiltonian in odd dimensions}

\ShortTitle{TKNN for general Hamiltonian}

\author{
Hidenori Fukaya\\
Department of Physics, Osaka University,
 Toyonaka, Osaka 560-0043, Japan.\\
hfukaya@phys.sci.osaka-u.ac.jp
 }
 
\author{
\speaker{
Tetsuya Onogi}\\
Department of Physics, Osaka University,
 Toyonaka, Osaka 560-0043, Japan.\\
onogi@phys.sci.osaka-u.ac.jp
}

 \author{
 Satoshi Yamaguchi\\
Department of Physics, Osaka University,
 Toyonaka, Osaka 560-0043, Japan\.\
 yamaguch@het.phys.sci.osaka-u.ac.jp
 }

 \author{
 Xi Wu\\
Physics Department, Ariel University, Ariel 40700, Israel\\
wuxi5949@gmail.com
}

\abstract{Topological insulators in odd dimensions are characterized by topological numbers. We prove the well-known relation between the topological number given by the Chern character of the Berry curvature and the Chern-Simons level of the low energy effective action for a general class of Hamiltonians bilinear in the fermion with general U(1) gauge interactions including non-minimal couplings by an explicit calculation. A series of Ward-Takahashi identities are crucial to relate the Chern-Simons level to a winding number, which could then be directly reduced to Chern character of Berry curvature by carrying out the integral over the temporal momenta.}

\FullConference{37th International Symposium on Lattice Field Theory - Lattice2019\\
		16-22 June 2019\\
		Wuhan, China}

\begin{document}

\section{Introduction}

Topological insulators in $D=2n+1$ dimensions are characterized either by the Chern character of the Berry connection from the eigenfunctions of the Hamiltonian in the valence band~\cite{Thouless:1982aa, Niu:1985aa} or  by the coefficient of the effective action as a functino of an external U(1) gauge field, i.e., photon~\cite{Ishikawa:1983ad,Ishikawa:1986wx,So:1984nf,Golterman:1992ub}.
 These two characterization are known to be equivalent because they both arise from the current correlation functions and there are explicit proofs for various cases \cite{Witten:2015aoa, Zubkov:2016mvv}.
In Ref.~\cite{Qi:2008ew} a proof is given for a large class of models for general odd dimensions, where they consider the most general lattice action for arbitrary free kinetic term on the lattice which is then coupled to U(1) gauge field in a minimal way, i.e. with the gauge interaction in the form of 
\begin{eqnarray}
H(A) = \sum_{m,n} \psi^\dagger_m h_{mn} e^{iA_{mn}} \psi_n + \sum_m \psi_m^\dagger \psi_m,
\end{eqnarray}
where $m,n$ are the lattice sites $h_{mn}$ are the hopping parameters and $A_{mn}$ is the line integral of the gauge field along the straight line connecting the sites $m$ and $n$.  The advantage of this class of Hamiltonian is that the contact interactions such as fermion-fermion-multi-photon vertices do not contribute to the final expression so that only a set of Feynman diagram which appear  also in the continuum theory gives non-vanishing contributions. Of course, this type of gauge interaction is physically motivated since it is based on the famous method of `Peierls substitution'~\cite{Peierls:1933}. However,  in more general situation, the gauge interaction may not always be described by such a single straight Wilson-line.  In such cases, one has to include the contribution of contact interaction vertices. 
In this proceedings, we report our recent study on the equivalence for generalized lattice Hamiltonian so as to include arbitrary non-minimal gauge interactions~\cite{Fukaya:2019urq}.

\section{Gapped fermion system on the lattice}

We consider a gapped fermion system (with a gap $\Delta$) on a lattice  with the following action in Euclidean space 
in $D=2n+1$ dimensions. 
\begin{eqnarray}
S_{\rm E} = \int dt \sum_{\vec{r}}\psi^\dagger(t,\vec{r}) 
\left[\frac{\partial}{\partial t}  + i A_0 + H(\vec{A})\right] \psi(t,\vec{r}),
\end{eqnarray}
where $\vec{r}$ runs over the $2n$ dimensional spatial lattice points. 
We will set $x^0=t$ in the followings. 
The Hamiltonian $H(\vec{A})$ is given by a summation over all the possible hoppings on the lattice 
which include
gauge interactions with a smooth external U(1) gauge field $A_\mu=(A_0,\vec{A})$. The fermion fields $\psi^\dagger(t,\vec{r})$ and $\psi(t,\vec{r})$ give creation and annihilation operators of  fermions after quantization. We assume that there are $N_v$ bands and $N_c$ bands below and above the fermi level, respectively.  Therefore the fermion fields have $N_v+N_c$ components. 

Since the fermion system is gapped with a gap size 
$\Delta >0$
, the effective gauge action 
obtained by
integrating out fermions can be expanded in terms of gauge invariant local actions as
$S_{\rm eff}= \sum_{k} a_k S_k(A)$.
Here, $S_{\rm eff}(A)$ is defined as 
$e^{S_{\rm eff}} = \int \mathcal{D}\psi\mathcal{D}\psi^\dagger e^{-S_{\rm E}}$,
and $S_k(A)$ are the gauge invariant actions given by the local Lagrangian $L_k(A)$ and $a_k$ are
the coefficients. By dimensional analysis, if the Lagragian $L_k(A)$ has a mass dimension $d_k$
the coefficient $a_k$ is suppressed by the $d_k-(2n+1)$-powers in $\frac{1}{\Delta}$ or lattice spacing $a$. Many of the Lagrangians are given in terms of gauge invariant field $F_{\mu\nu}$
Since we do not have the Lorentz-invariance on the lattice,  the structure of the coefficients $a_k$ in the effective action heavily depend the geometry of the lattice. 
However, there is a very special parity-violating term called Chern-Simons action $S_{\rm cs}(A)$ given by
\begin{eqnarray}
S_{\rm cs}(A) = \int d^{2n+1} x ~ \epsilon_{\alpha_0 \beta_1\alpha_1 \cdots \beta_{n}\alpha_{n}}
A_{\alpha_0} \partial_{\beta_1} A_{\alpha_1} \cdots 
\partial_{\beta_{n}} A_{\alpha_{n}} .
\end{eqnarray}
This action is topological and always takes this form no matter what the geometry of the lattice is.
Topological information of the fermion system is contained in the effective action through the coefficient $c_{\rm cs}$ as
$S_{\rm eff}(A) =  i c_{\rm cs} S_{\rm cs} + \cdots$ where "$\cdots$ " stands for other gauge invariant terms .
Here the gauge invariance of the action requires that the coefficient is quantized as
\begin{eqnarray}
c_{\rm cs} = \frac{k}{(2\pi)^n (n+1)!}  &, ~~& k\in \mathbb{Z} .
\label{eq:ccs_k}
\end{eqnarray}
Since the Chern-Simons action is of the lowest dimension in the parity-violating sector, the coefficient $c_{\rm cs}$ can be obtained by the following  quantity
\begin{eqnarray}
c_{\rm cs}
&=&
\frac{(-i)^{n+1} \epsilon_{\alpha_0\beta_1 \alpha_1\cdots \beta_n \alpha_n}}{(n+1)!(2n+1)!}
\left(\frac{\partial}{\partial (q_1)_{\beta_1}}\right)
\cdots
\left(\frac{\partial}{\partial (q_n)_{\beta_n}}\right)
\left.
\prod_{i=1}^n\int d^D x_i
e^{iq_{\alpha_i} x_i}
\frac{\delta^{n+1} S_{\rm eff}(A)}
{\delta A_{\alpha_0}(x_0) \delta  A_{\alpha_1}(x_1)
\cdots \delta A_{\alpha_n}(x_n)}
\right|_{q_i=0} .
\label{eq:Ccs_Seff}
\end{eqnarray}



The effective action can be given by the log of the fermion determinant as
$S_{\rm eff}(A) = \mbox{Tr}\left[\ln\left(D_0 + H(A)\right)\right]$,
where $D_0= \frac{\partial}{\partial x^0}+i A_0$.
Splitting the kinetic operator $D_0 + H(A)$ into free part and interaction part as
$D_0 + H(A) = \frac{\partial}{\partial x_0} + H_0 - \Gamma(A$), 
where $H_0$ is the free fermion part defined as $H_0\equiv \left. H(A)\right|_{A=0}$ and $\Gamma(A)$ is the interaction part defined as 
$\Gamma(A) \equiv - iA_0 - H(A) + H_0$.
Expanding in $\Gamma(A)$,  we obtain
\begin{eqnarray}
S_{\rm eff}(A) -\mbox{\rm const.}&=& - \sum_{n=1}^\infty \frac{1}{n}\mbox{Tr}\left[\left(\frac{1}{\frac{\partial}{\partial x_0} + H_0}\Gamma(A)\right)^n\right] 
\end{eqnarray}

From Eq.(\ref{eq:Ccs_Seff}), we find that the Chern-Simons 
coupling
for $D=2+1$ dimension is given by
\begin{eqnarray}
c_{\rm cs} 
&=& 
-\frac{(-i)^2}{2!3!}\int \frac{d^3p}{(2\pi)^3}
\epsilon_{\alpha_0\beta_1\alpha_1}\left(\frac{\partial}{\partial q_1}\right)_{\beta_1}
\nonumber\\
&&
\left\{
\mbox{Tr}\left[S_F(p) 
\Gamma^{(2)}[-q_1,\alpha_0;q_1,\alpha_1;p]\right]\right.
 +
\left.
\left.\mbox{Tr}\left[S_F(p-q_1) 
\Gamma^{(1)}[-q_1,\alpha_0;p]
 S_F(p) 
  \Gamma^{(1)}[q_1,\alpha_1;p-q_1] \right]
\right\}\right|_{q_1=0},
\label{eq:ccs_3D}
\end{eqnarray}
where $S_F(p)$ is the fermion propagator $\frac{1}{ip_0 +H_0(\vec{p})}$ and $\Gamma^{(1)}[q_1,\alpha_1;p]$ and $\Gamma^{(2)}[q_1,\alpha_1;q_2,\alpha_2;p]$ 
are fermion-fermion-photon and fermion-fermion-photon-photon vertices  with in-coming fermion momentum $p$ and in-coming photon momenta $q_i ~(i=1,2)$ with Lorentz index 
$\alpha_i ~(i=1,2)$
\begin{eqnarray}
\Gamma^{(1)}[q_1,\alpha_1;p]
&=& \int d^{2n+1} x_1 ~e^{iq_1\cdot x_1}\int d^{2n+1} y ~ e^{ip \cdot y}
\left. \frac{\delta \Gamma[A](x,y)}{\delta A_{\alpha_1}(x_1)} \right|_{A=0},
\\
\Gamma^{(2)}[q_1,\alpha_1,q_2;\alpha_2;p]
&=& \prod_{i=1}^2\left(\int d^{2n+1} x_i~ e^{iq_i\cdot x_i}\right)\int d^{2n+1} y ~ e^{ip \cdot y}
\left. \frac{\delta^2 \Gamma[A](x,y)}{\delta A_{\alpha_1}(x_1)\delta A_{\alpha_2}(x_2)} \right|_{A=0},
\end{eqnarray}
Note that the contributions with multi-photon vertices vanishes for the class of Hamiltonians with gauge interactions given by a single straight Wilson-line because the multi-photon vertices are symmetric under the interchange of Lorentz indices of photons. When contracted with the antisymmetric tensor, such contributions vanish. However, in general Hamiltonian we must consider these contributions.

In  addition to the usual  Ward-Takahashi identity $\Gamma^{(1)}[0,\alpha;p] = \textcolor{blue}{-}\frac{\partial S^{-1}_F(p)}{\partial p_\alpha}$,
we derive 
the Ward-Takahashi identities in Appendix~ \ref{App_WT} as 
\begin{eqnarray}
\left.\frac{\partial^2 \Gamma^{(1)}[k,\mu;p]}{\partial k_\nu \partial p_\lambda}\right|_{k=0}
=\left. \frac{\partial \Gamma^{(2)}[k,\mu;0,\lambda;p]}{\partial k_\nu}\right|_{k=0}
=\left. \frac{\partial \Gamma^{(2)}[0,\lambda;l,\mu;p]}{\partial l_\nu}\right|_{l=0} .
\label{eq:2nd_WT}
\end{eqnarray}
Using these identities,  the integrand of Eq.~(\ref{eq:ccs_3D}) can be rewritten as 
\begin{eqnarray}
\epsilon_{\alpha_0\beta_1 \alpha_1}
\mbox{Tr}
\left[
\frac{\partial}{\partial_{p_{\alpha_0}}}
\left(2S_F(p)
\frac{\partial \Gamma^{(1)}[q, \alpha_1;p]}{\partial q_{\beta_1} }
\right)
\left.
+
S_F(p)
\frac{\partial S^{-1}_F(p)}{\partial p_{\alpha_1}}
S_F(p)
\frac{\partial S^{-1}_F(p)}{\partial p_{\beta_1}}
S_F(p)
\frac{\partial S^{-1}_F(p)}{\partial p_{\alpha_0}}
\right]
 \right|_{q=0}
.
\label{eq:winding_3D}
\end{eqnarray}
The first  term is a total divergence which  vanishes when we integrate over the momentum. Therefore, one finds that the Chern-Simons 
coupling
is given by the winding number as
\begin{eqnarray}
c_{\rm cs}
&=&
\frac{(-i)^2 \epsilon_{\alpha_0\beta_1 \alpha_1}}{2!3!}
\int\frac{dp_0}{2\pi}
\int_{\rm BZ}
\frac{d^{2}p}{(2\pi)^{2}}
\mbox{Tr}
\left[
S_F(p)
\frac{\partial S_F^{-1}(p)}{\partial p_{\alpha_0}}
S_F(p)\frac{\partial S^{-1}_F(p)}{\partial p_{\beta_1}}
S_F(p)\frac{\partial S_F^{-1}(p)}{\partial p_{\alpha_1}}
\right].
\end{eqnarray}
For the case for $D=4+1$ dimension~($n=2$), we also have 
the following Ward-Takahashi identitiy given in Appendix\ref{App_WT} 
\begin{eqnarray}
\left.\frac{\partial^2 \Gamma^{(3)}[q,\mu; r,\nu;0,\lambda;p]}{\partial q_{\alpha} \partial r_{\beta}}\right|_{q=r=0}
= \left.\frac{\partial^3\Gamma^{(2)}[q,\mu; r,\nu;p]}{\partial q_{\alpha} \partial r_{\beta} \partial p_\lambda}\right|_{q=r=0}.
\label{eq:WT_G3}
\end{eqnarray}
with which one can show extra terms add up to total derivatives so  we obtain 
\begin{eqnarray}
c_{\rm cs} 
&=& 
-
\frac{(-i)^3\cdot 2}{3!5!}\int \frac{d^5p}{(2\pi)^5}
\epsilon_{\alpha_0\beta_1\alpha_1\beta_2\alpha_2} 
\nonumber\\
&&
\mbox{Tr}
\left[
S_F(p)
\frac{\partial S^{-1}_F(p)}{\partial p_{\alpha_0}}
 S_F(p)
\frac{\partial S^{-1}_F(p)}{\partial p_{\beta_1}}
 S_F(p)
\frac{\partial S^{-1}_F(p)}{\partial p_{\alpha_1}}
 S_F(p)
\frac{\partial S^{-1}_F(p)}{\partial p_{\beta_2}}
 S_F(p)
\frac{\partial S^{-1}_F(p)}{\partial p_{\alpha_2}}
\right].
\label{eq:winding_5D}
\end{eqnarray}
Therefore, Chern-Simons coupling is given by the winding number with fermion propagator also for $D=4+1$ case.

\section{Equivalence of  winding number and chern number}
We now show the equivalence of the Chern-Simons coupling given by the winding number expression and the Chern character given by the Berry connection 
for the energy eigenstates in the valence bands. 
The proof of this part is already given in Ref.~\cite{Qi:2008ew}, but since the proof is simple, we give it here for completeness.
We give the calculation for arbitrary odd ($D=2n+1$) dimensions, even though we have shown that the Chern-Simons coupling $c_{\rm cs}$  
can be written by the winding number using $S_F$ only for $D=2+1$ and $D=4+1$ dimensions.

In order to simplify the notation, hereafter we abbreviate the derivative with respect to the momentum $p_\mu$ 
as $\partial_\mu \equiv \frac{\partial}{\partial p_\mu}$.
The result of the previous section for $D=2+1$ and $D=4+1$ can be unified to the following results:
In the expression using the fermion propagator $S(p)=\frac{1}{ip^0+H}$ and inserting a complete set  of energy eigenstates $\displaystyle{\sum_{\alpha}|\alpha\rangle\langle \alpha |}$, where $\alpha$ is the label of energ, the Chern-Simons coupling $c_{\rm cs}$ is  given as
\begin{eqnarray}
	c_{\rm cs}
	&=&\frac{n! (-i)^{n+2}}{(n+1)!(2n)!}\int \frac{d^{2n}p}{(2\pi)^{2n}} 
	\sum_{\alpha_1,\cdots, \alpha_{2n}} \epsilon^{i_1 i_2 \cdots i_{2n}}
\int \frac{dp^0}{2\pi}
	\frac{\langle \alpha_1|\partial_{i_1} H |\alpha_2\rangle\langle \alpha_2 |\partial_{i_2} H | \alpha_3\rangle 
                \cdots \langle \alpha_{2n} |\partial_{i_{2n}} H | \alpha_1\rangle}
	{(ip^0+E_{\alpha_1})^2(ip^0+E_{\alpha_2})\cdots(ip^0+E_{\alpha_{2n}})},
 .
\label{eq:ccs_J2n+1}
\nonumber\\
\end{eqnarray}
where $i_1,\cdots, i_{2n}$ stand for the spatial indices. 

All we have to do is to integrate over $p^0$ using Cauchy's theorem. 
Here, we use a trick to simplify the integration. It is easy to see that the expression Eqs.~(\ref{eq:winding_3D}), (\ref{eq:winding_5D}) are invariant under continuous deformation of $S_F$ (or $H$) provided that the integrand remains to have no singularities. Therefore,  under  a continuous change of the Hamilitonian, the winding number remains unchanged from its original value as long as the enegry spectrum is kept gapped throughout the deformation. 
Now, the most general Hamiltonian with $N_v$ valence bands and $N_c$ conduction bands is expressed as
\begin{eqnarray}
H(\vec{p}) \equiv \sum_{a=1}^{N_v} E_a(\vec{p})  |a(\vec{p})\rangle \langle a(\vec{p}) |
+\sum_{\dot{b}=1}^{N_c} E_{\dot{b}}(\vec{p})  |\dot{b}(\vec{p})\rangle \langle \dot{b}(\vec{p}) |,
\end{eqnarray}
where $|a(\vec{p})\rangle$ labeled by $a$ is the energy eigenstate in the valence band with spatial momentum $\vec{p}$ and negative energy eigenvalue $E_a(\vec{p})<0$. 
The state $|\dot{b}(\vec{p})\rangle$ labeled by $\dot{b}$ is the energy eigenstate in the conduction band with spatial momentum $\vec{p}$ and positive energy eigenvalue $E_{\dot{b}}(\vec{p})>0 $. One can continuously deform the Hamiltonian without hitting the singularity of $S(p)$ ( i.e. keeping the system gapped ) so that all energy eigenvalues in the conduction bands and all energy eigenvalues in the valence bands are degenerate and momentum independent (i.e. flat band ) respectively. 
Then the deformed Hamiltonian $H_{\rm new}$ which gives the same winding number becomes
\begin{eqnarray}
H_{\rm new}(\vec{p}) = 
E_v \sum_{a=1}^{N_v}  |a(\vec{p})\rangle \langle a(\vec{p}) |
+ E_c \sum_{\dot{b}=1}^{N_c}   |\dot{b}(\vec{p})\rangle \langle \dot{b}(\vec{p}) |,
\end{eqnarray}
where $E_v < 0$, $E_c>0$ are the momentum independent constant.  Here the eigenstates are identical to those with the original Hamiltonian.
Since there are only two poles, we can easily carrry our $p^0$ integral to obtain
\begin{eqnarray}
J = - \sum_{a_1,\cdots,a_n=1}^{N_v} \sum_{\dot{a}_1,\cdots,\dot{a}_n=1}^{N_c}  \epsilon^{i_1 j_1 \cdots i_{2n} j_{2n}} 
 \frac{(2n)!}{(n!)^2}
\langle a_1 | \partial_{i_1} \dot{a}_1\rangle \langle \dot{a}_1 | \partial_{j_1} a_2\rangle
\times \cdots  \times    \langle a_n|\partial_{i_n} \dot{a}_n\rangle\langle \dot{a}_n|\partial_{j_n} a_1\rangle .
\nonumber
\end{eqnarray}
Let us define the Berry connection using the negative energy eigenstates as
\begin{eqnarray}
\mathcal{A}^{ab} \equiv \mathcal{A}^{ab}_\mu dx^\mu = - i \langle a | \partial_\mu b \rangle dx^\mu
\equiv -i \langle a | db\rangle .
\end{eqnarray}
Then it is straightforward to show that the Berry curvature $\mathcal{F}^{ab}$ is 
\begin{eqnarray}
\mathcal{F}^{ab} &\equiv&\left( d\mathcal{A} + i \mathcal{A} \mathcal{A} \right)^{ab}
=       i \sum_{\dot{c}=1}^{N_c} \langle a| d\dot{c}\rangle \langle \dot{c} | d b\rangle 
\label{eq:Berry_curvature}
\end{eqnarray}
Using Eq.~(\ref{eq:Berry_curvature}), the Chern-Simons coupling can be expressed as 
\begin{eqnarray}
c_{\rm cs}
=  \frac{(-1)^n}{(n+1)!(2\pi)^n} \int_{BZ} \mbox{ch}_n(\mathcal{A}),
\end{eqnarray}
where $\mbox{ch}_n(\mathcal{A})$ is the 2nd Chern character defined by
$\mbox{ch}_n(\mathcal{A})=\frac{1}{n!}  \frac{1}{(2\pi)^n}
 \mbox{tr}( \mathcal{F}^n )$ . 
Comparing this expression with Eq.~(\ref{eq:ccs_k}) 
$c_{\rm cs} = \frac{k}{(n+1)!(2\pi)^n}$,
\begin{eqnarray}
k= (-1)^n \int_{BZ}  \mbox{ch}_n(\mathcal{A}).
\end{eqnarray}
We have shown  the Chern-Simons level and the topological number in terms of the Berry connection is  identical. On the other hand, while the Berry connection approach is limited to the free theory case, the effective theory approach can be applied also to interacting theories. 

\begin{acknowledgments}
 This work is supported in part by the Japanese Grant-in-Aid for Scientific Research(Nos. 15K05054, 18H01216, 18H04484 and 18K03620).
\end{acknowledgments}

%
%
%
\appendix

\section{Ward-Takahashi identities}
\label{App_WT}
In this appendix, we derive various identities among vertex functions and the inverse fermion propagator obtained from gauge invariance, i.e. Ward-Takahashi identities.
Finite difference operator which appears in the hopping term of the lattice fermion system can be expressed in terms of infinite series of derivatives. Therefore, we assume that the Hamiltonian can be expressed in terms of all sorts of fermion hopping terms 
connected by the Wilson-lines of arbitrary contours or superpositions of them.
Then, the action can be formally expanded as
\begin{eqnarray}
S = \int dt \sum_{\vec{x}} \sum_{n=0}^\infty \psi^\dagger(t,\vec{x})
M_{\mu_1 \cdots \mu_n} (D_{\mu_1} \cdots D_{\mu_n}\psi)(t,\vec{x})
\label{eq:action}
\end{eqnarray}
where summation over $\mu_1, \cdots,\mu_n$ are implicit. $M_{\mu_1\cdots\mu_n}$ are some $N\times N$ matrix where $N=N_c+N_v$ is the number of fermion degrees of freedom per site.

Expanding this action in terms of gauge fields and making Fourier transformations, one can obtain the formal expressions of the inverse propagator and the vertex functions in the momentum space.
In the following, let us denote the inverse fermion propagator with momentum $p$ as $S^{-1}_F(p)$ and the vertex functions with incoming fermion momentum $p$ and $n$ photons with incoming momentum $k_i$ and $\mu_i$ components ($i=1,\cdots, n$) and outgoing fermion with momentum $p+ \displaystyle{\sum_{i=1}^n k_i}$ as $\Gamma^{(n)}[k_1,\mu_1;\cdots;k_n,\mu_n;p]$. Then the formal expression gives
\begin{eqnarray}
&&S_F^{-1}(p) = \sum_{n=0}^\infty M_{\mu_1\cdots\mu_n} \prod_{i=1}^n \left(ip_{\mu_i}\right),
\label{eq:S_F}
\end{eqnarray}
\begin{eqnarray}
&&\Gamma^{(1)}[k,\mu;p] =
-
i \sum_{n=1}^\infty \sum_{a=1}^nM_{\mu_1\cdots\mu_{a-1} \mu\mu_{a+1}\cdots \mu_n} \prod_{i=1}^{a-1} \left(i(p+k)_{\mu_i}\right) \prod_{i=a+1}^{n} \left(ip_{\mu_i}\right),
\label{eq:Gamma^{(1)}}
\end{eqnarray}
Substituting Eqs. (\ref{eq:S_F}) (\ref{eq:Gamma^{(1)}}), it is easy to show the usual Ward-Takahashi identity in QED holds.

Now, t is interesting to note that we could also obtain Ward-Takahashi identities for quantities involving higher order terms in photon momenta and multi-photon vertex functions. 
For example the two-photon vertex is given as 
\begin{eqnarray}
&&\Gamma^{(2)}[k,\mu;l,\nu;p] 
\nonumber\\
&=& 
-
i^2\sum_{n=1}^\infty 
\sum_{\underset{a<b}{a,b=1}}^{n}
M_{\mu_1\cdots\mu_{a-1} \mu\mu_{a+1}\cdots \mu_{b-1}\nu\mu_{b+1}\cdots\mu_n} \prod_{i=1}^{a-1} \left(i(p+k+l)_{\mu_i}\right) \prod_{i=a+1}^{b-1} \left(i(p+l)_{\mu_i}\right) \prod_{i=b+1}^{n} \left(ip_{\mu_i}\right)
\nonumber\\
&
-
&i^2\sum_{n=1}^\infty 
\sum_{\underset{a<b}{a,b=1}}^{n}
M_{\mu_1\cdots\mu_{a-1} \nu\mu_{a+1}\cdots \mu_{b-1}\mu\mu_{b+1}\cdots\mu_n} \prod_{i=1}^{a-1} \left(i(p+k+l)_{\mu_i}\right) \prod_{i=a+1}^{b-1} \left(i(p+k)_{\mu_i}\right) \prod_{i=b+1}^{n} \left(ip_{\mu_i}\right),
\nonumber\\
\label{eq:Gamma^{(2)}}
\end{eqnarray}
Using Eqs.(\ref{eq:Gamma^{(1)}}) (\ref{eq:Gamma^{(2)}}) , we obtain the following identities:
\begin{eqnarray}
\left.\frac{\partial^2 \Gamma^{(1)}[k,\mu;p]}{\partial k_\nu \partial p_\lambda}\right|_{k=0}
=\left. \frac{\partial \Gamma^{(2)}[k,\mu;l,\lambda;p]}{\partial k_\nu}\right|_{k,l=0}
=\left. \frac{\partial \Gamma^{(2)}[k,\lambda;l,\mu;p]}{\partial l_\nu}\right|_{k,l=0}
\end{eqnarray}
Carrying out similar calculations by simply differentiaing $\Gamma^{(2)}$ and $\Gamma^{(3)}$ , 
we can see that the following identity holds:
\begin{eqnarray}
\left.\frac{\partial^2 \Gamma^{(3)}[q,\mu; r,\nu;s,\lambda;p]}{\partial q_{\alpha} \partial r_{\beta}}\right|_{q,r,s=0}
= \left.\frac{\partial^3\Gamma^{(2)}[q,\mu; r,\nu;p]}{\partial q_{\alpha} \partial r_{\beta} \partial p_\lambda}\right|_{q,r=0}
\end{eqnarray}





%
%

%

\end{document}